\documentclass[prb,twocolumn,amsmath,amssymb,showpacs,letterpaper]{revtex4-1}
\usepackage{graphicx}
\usepackage{verbatim}
\begin{document}
\title{RTSPM: Real-time Linux control software for scanning probe microscopy}

\author{V. Chandrasekhar}
\email{v-chandrasekhar@northwestern.edu}
\author{M. M. Mehta}
\affiliation{Department of Physics and Astronomy, Northwestern University, Evanston, Illinois. 60208, USA}


\pacs{07.79.-v, 89.20.Ff}

\begin{abstract}
Real time computer control is an essential feature of scanning probe microscopes, which have become important tools for the characterization and investigation of nanometer scale samples.  Most commercial (and some open-source) scanning probe data acquisition software uses digital signal processors (DSPs) to handle the real time data processing and control, which adds to the expense and complexity of the control software.  We describe here scan control software that uses a single computer and a data acquisition card to acquire scan data.  The computer runs an open-source real time Linux kernel, which permits fast acquisition and control while maintaining a responsive graphical user interface. Images from a simulated tuning-fork based microscope as well as a standard topographical sample are also presented, showing some of the capabilities of the software.   
\end{abstract}

\maketitle

Scanning probe microscopy (SPM) in all its forms has developed into a powerful tool for the study of the nanometer scale properties of materials and devices.\cite{kalinin} Computer control has been an essential part of SPM since its inception:  it is required both for the control of hardware, and the acquisition and processing of data.  It would not be an exaggeration to say that the proliferation of scanning probe microscopes as a common tool in research laboratories has been driven by the availability of cheap computing power.  In order to obtain an image in a reasonable amount of time, hardware control must be performed with very short response times.  In earlier incarnations of scanning probe microscopy, many of the time-critical functions like feedback control loops were performed with analog electronics.  However, there is a significant advantage of flexibility in all-digital control, and the trend in commercial software has been increasingly towards full digital control.

Most modern computers run multitasking operating systems (OSs), in that they nominally handle many different processes and tasks simultaneously.  In reality, however, a single CPU may run only one or a few tasks at one time, so that the many different threads and processes that run concurrently in the operating system are scheduled to run for a given time slice before being switched out and having to await their turn again.  On a sufficiently powerful computer, this happens fast enough that it appears a number of programs are running in parallel.  For example, on a Microsoft Windows computer, a particular process will be run on the CPU on average every 55 ms (typically 10 ms on a more recent computer).  Linux has similar time constraints.  While this may be fast enough for programs that interact with human users, it is not fast enough for the hardware control loops  required for scanning probe microscopy, which typically require response times of 1 ms or better.

This limitation has been overcome in the past by employing analog control electronics for fast control.  More recently, for full digital control, the usual technique is to place a dedicated real time digital controller between the computer and the hardware.\cite{zahl}  This is usually a digital signal processor (DSP) or a field programmable gate array (FPGA) that is interfaced to digital-to-analog (DAC) and analog-to-digital (ADC) cards or chips which ultimately communicate with the microscope itself.  DSPs and FPGAs can match the time response of analog electronics while providing the flexibility of a digitally programmable tool.  However, DSPs and FPGAs require specialized software tools and significant time and effort to program, and thus may not be accessible to those on limited budgets.

We describe here an open-source scanning probe control program \cite{github} that runs on a standard personal computer with a relatively inexpensive data acquisition card.  The computer runs a custom Linux kernel patched with the Real Time Application Interface (RTAI).\cite{rtai}  While the minimum response times (25 $\mu$s) are not as short as those obtainable with a modern DSP or FPGA controller, they are short enough for controlling the feedback loops required for scanning probe microscopy.  In addition, all the software required to compile the kernel and control the data acquisition card is open-source and freely available, as are the Integrated Development Environments (IDEs) used to program the software. 

\section{Software and hardware description}
\subsection{Computer and software}
The core of our scanning probe control system is a computer with a data acquisition card running a custom compiled Linux kernel patched with the Real Time Application Interface.\cite{rtai}  Almost any recent computer will work, although as with almost all computers, purchasing the most powerful computer that the budget allows is advisable.  In particular, it is useful to have a multicore CPU, as real-time processes can be directed to run on specific cores.  For the work described here, we have used a Dell T110 server with a Xeon 2.4 GHz 4-core processor and 4 GB of RAM, as well as a home-assembled computer with an Intel Core2 dual core computer running at 3.06 GHz with 2 GB of memory for the tuning-fork simulator (details of the tuning fork simulator will be discussed in another publication).  
\begin{figure}
\includegraphics[width=8.0cm]{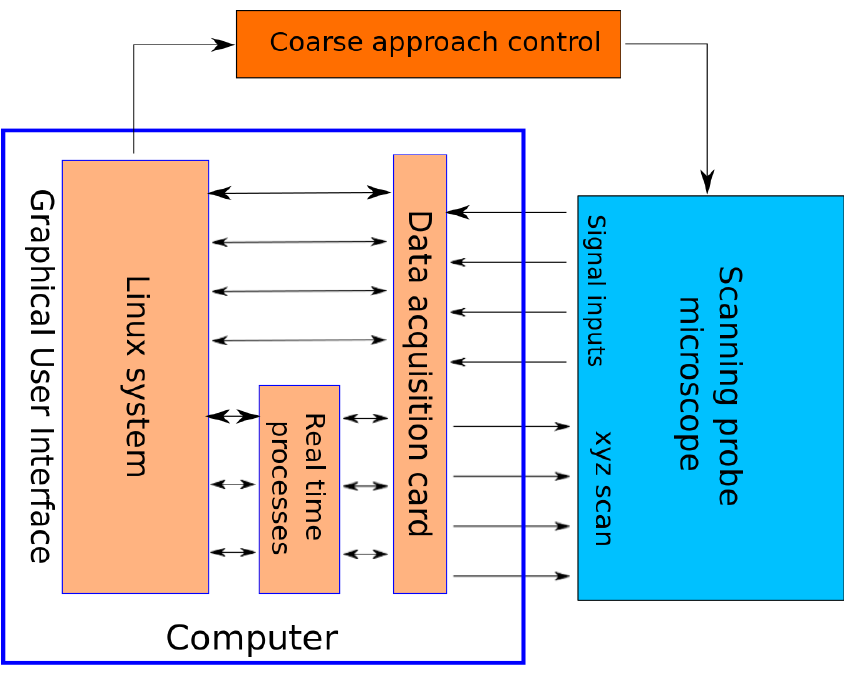}
\caption{Schematic of the real-time computer control system.}
\label{fig1}
\end{figure}
There are a number of options for real-time Linux kernels, including RTAI, Xenomai,\cite{xenomai} RTOS\cite{rtos} and others.  We chose to go with RTAI as its integration with the open source data acquisition driver software Comedi\cite{comedi} allowed for support of the largest number of data acquisition boards.  Here we use data acquisition boards from National Instruments (\url{http://www.ni.com}), but boards from a number of other manufacturers are also supported.

Patching and installing a Linux kernel with real-time extensions, although sometimes tricky, is well documented,\cite{rtai} so we shall not cover it here.  However, since we require the Comedi extensions, we recommend that instructions for installing RTAI-Lab be followed.\cite{rtailab}  This installs a kernel with suitable real-time extensions, the Comedi drivers and Comedi real-time extensions.  It also enables one to install ScicosLab,\cite{scicoslab} a graphical programming environment that can run real-time tasks, so that the installation of the real-time kernels, drivers and performance can be tested.  Note that RTAI, Comedi and ScicosLab are available and can be installed as packages on many Linux distributions, including the Debian-based Ubuntu distribution that we used, but for the required integration, it is necessary to compile the software from source.

To write the real-time programs, a standard C/C++ compiler is required: this usually comes with the Linux distribution, or can be installed easily from the software repositories.  Any text editor can be used to write the programs, which can then be compiled from a command line.  We chose instead to use the open-source cross-platform C/C++ IDE Code::Blocks,\cite{codeblocks} which provided a convenient environment in which to write, compile and debug programs.  This is also available as a pre-compiled package on many Linux software repositories.  

Figure 1 shows a schematic of the scanning microscopy setup.  The RTAI patch to the Linux kernel makes available an Applications Programming Interface (API) that gives priority to programmed real-time tasks, so that the usual Linux kernel is run as a background process.  Two types of real-time programs can be compiled:  the so-called kernel space and user space programs.  Kernel space programs are compiled as modules that can be loaded into the Linux kernel.  They are generally slightly faster than user space programs; however, it is easier to share data between a user space program and a regular Linux programs.  Indeed, our real-time programs are written as functions and subroutines that be can started and stopped from the main Linux GUI program, and parameters can be changed on the fly from within the GUI without stopping the real-time processes. 

A critical advantage of the real-time interface for scanning probe microscopy is the possibility to execute timed loops with well-determined time intervals in the microsecond range.  To output a simple sine wave, which involves writing two bytes to the DAC every cycle, we can achieve deterministic loop times of 5 $\mu$s.  (This time is determined by the interrupt handling routines of the BIOS and kernel.)  However, because the computer is spending so much time servicing the real time requests, this freezes the GUI.  Consequently, for typical proportional-integral-differential (PID) control loops, where a significant amount of calculation may also be required, a minimum time of 25 $\mu$s is necessary.  In order to avoid any overload issues, we typically use more conservative loop times of 50 to 100 $\mu$s for our control loops, which is fast enough for our purposes. 

The real-time subroutines and programs communicate with the data acquisition card through the real-time Comedi extensions.  When the real-time program is reading data from a specific input channel, or writing data to a specific output channel, it locks access to the channel prior to data transfer, and unlocks it immediately after.  This means that the regular Linux GUI program can also communicate with all channels on the data acquisition card for non time-critical applications, for example during the close-approach process, when it extends the piezo in steps and checks for changes in the feedback signal at each step.  Indeed, all non real-time tasks such as coarse-approach and the actual $x-y$ scan are handled by the Linux GUI.

For programming the Linux GUI, one can use a number of open-source alternatives.  Of course, there is Code::Blocks itself, or Python,\cite{python} which has interfaces for both Comedi (pyComedi) and RTAI (python-rtai).  However, for the Linux GUI, our choice was to go with Free Pascal,\cite{freepascal} using the open-source Lazarus IDE \cite{lazarus} as the programming environment, as creating a GUI interface in this environment was much faster and easier, and also because Pascal is the preferred data acquisition programming language for one of us (VC).

\subsection{Hardware}
The hardware required for the RTSPM controller is a data acquisition card with a sufficient number of analog-to-digital (AD) inputs to read any required signals, and a sufficient number of digital-to-analog (DA) outputs to be able to control the required outputs, primarily the $x$, $y$ and $z$ piezo tube scanner voltages, and additional outputs if other parameters need to be controlled.  For these requirements, the speed constraints on the card are not severe:  if we need to update the output and read the input only every 50 $\mu$s, then a card with a 20 kHz update rate is sufficient.  Even the least expensive cards can easily achieve these rates, since they usually have update rates of at least 100 kHz.  For more sophisticated control, such as implementing a phase-locked-loop (PLL) from within the program, higher update rates would be required.  If a greater number of input and output channels are required, multiple cards can be used.  The program currently allows three cards to be used, but there 
is no 
reason why this cannot be increased.  This enables one to utilize older cards that individually may not have the required number of input and output channels.

Here we use a National Instruments (NI) PCIe-6259 DAQ card, which has 32 16-bit AD channels and 4 16-bit DA channels, giving us a sufficient number of outputs to control the $xyz$ axes of the piezo tube, and to simultaneously read multiple inputs, e.g., for topography, electrostatic force microscopy, etc.  The maximum ADC speed is 1.25 mega samples per second (MS/s), while the update rate for the DACs is 2.56 MS/s.  The system controls a home-made tuning-fork based scanning probe microscope that will be described in detail elsewhere: here we describe how it interacts with the scanning probe software.  Three DACs from the NI PCIe-6259 DAQ card control the $x$, $y$, and $z$ axes of a four quadrant piezo tube scanner through five independent home-made high voltage amplifiers ($\pm x$, $\pm y$ and $z$); the amplifiers have a gain of 15 so that $\pm 10$ V from the DACs gives $\pm 150$ V to the piezo tube.  An ac voltage is applied to the tuning fork with an attached tungsten tip at or near its mechanical 
resonance 
frequency, and the resulting current, which is amplified through a home-made current preamplifier, is a measure of the amplitude of oscillation.   The transduction of the force between the tip and the surface is achieved either by monitoring the shift in resonant frequency using a PLL (a NanoSurf easyPLL) (\url{http://www.nanoscience.com/}) or by monitoring the change in amplitude or phase at a fixed frequency using a lock-in amplifier (Signal Recovery 7260, \url{http://www.signalrecovery.com}). 

\begin{figure*}[ht!]
\vspace{0.25cm}
\includegraphics[width=15cm]{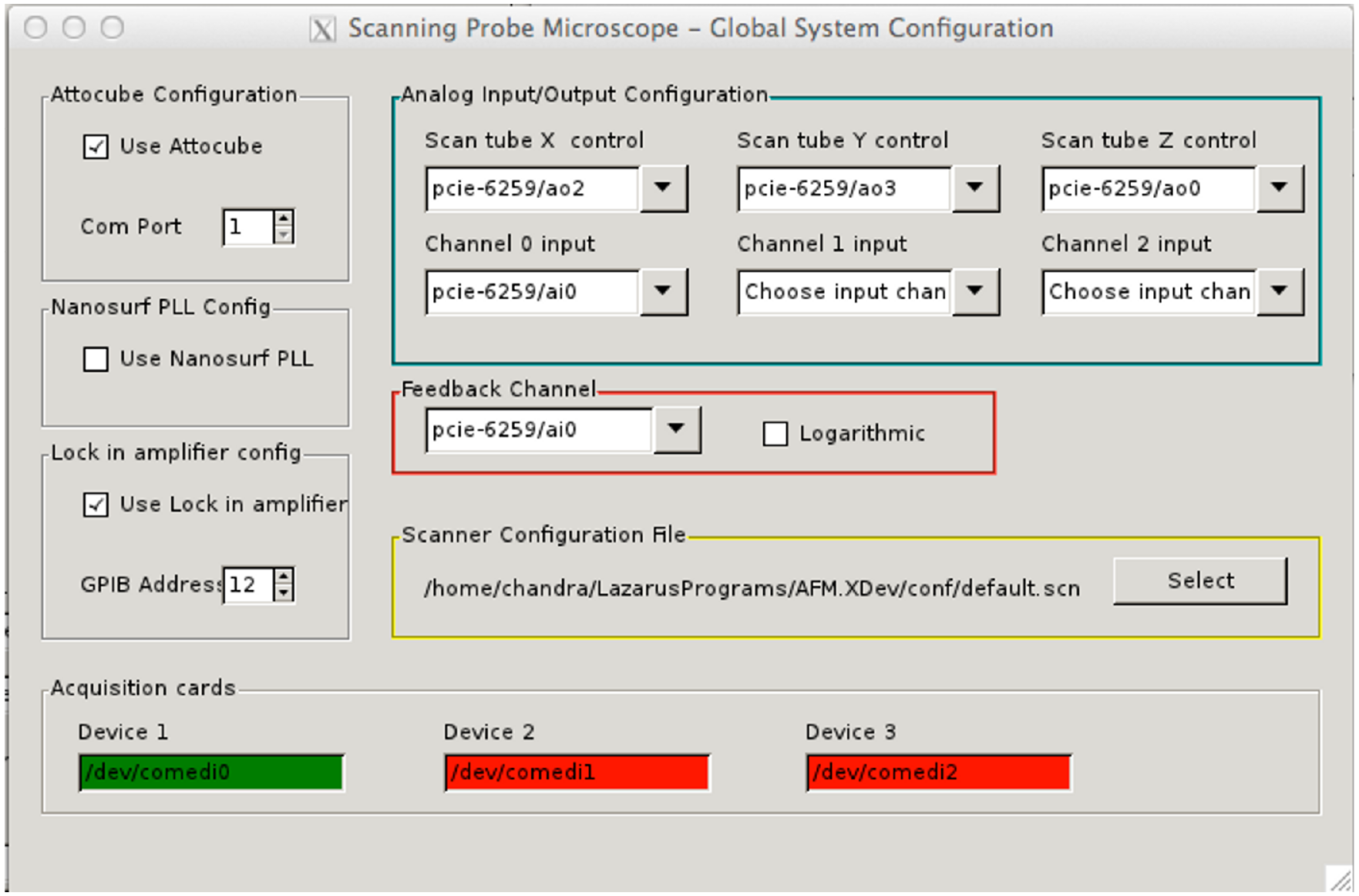}
\caption{System configuration panel}
\vspace{-0.25cm}
\label{fig2}
\end{figure*}
Coarse approach is achieved by controlling a $z$ coarse approach stage from Attocube (\url{http://www.attocube.com}) through their controller (ANC 150), which can be interfaced to the computer through a serial interface.  $x$ and $y$ axis coarse movement is achieved in the same manner.  The commercial Attocube coarse approach controller sends a simple sawtooth waveform to the coarse approach stages.  Consequently, coarse approach can also be controlled directly by the computer if additional analog outputs are available.  For example, on another system, we have used a NI PCI-6733 analog output card, which has 8 16-bit analog outputs, in tandem with a  NI PCI-6014 card, which in principle would allow us to do this. 

\section{RTSPM program}
\begin{figure}[h!]
\vspace{0.25cm}
\includegraphics[width=8.5cm]{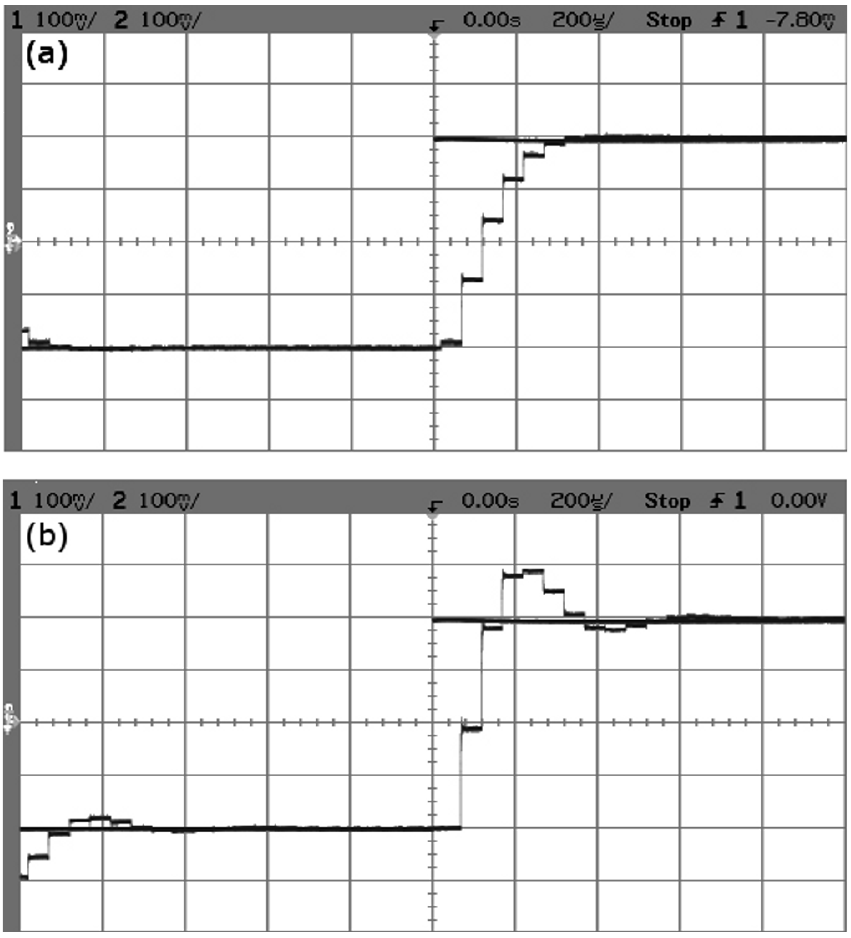}
\caption{(a)  Response of the real-time PID to a square wave error signal.  The PID parameters are proportional gain = 0.05, integral time constant 8 $\mu$s, and differential time constant 20 ns.  The PID response shows well-damped behavior.  (b)  PID response with proportional gain 0.08, integral time constant 8 $\mu$s and differential time constant 50 ns, demonstrating overshoot.  Each time division is 200 $\mu$s.}
\vspace{-0.5cm}
\label{fig3}
\end{figure}
\subsection{Overview}
The overall software design of the RTSPM program is similar to that found in many other programs, so we shall not discuss it in any detail here.  When the program is started, the user is presented with a dialog allowing the selection of a number of different options.  Initially, only the choice of the System Configuration Panel is allowed.    
Figure \ref{fig2}a shows a screenshot of the RTSPM configuration panel, where one defines the input and the output channels used by the program.  The program automatically detects the data acquisition cards as well as their A/D and D/A channels on the computer if they are recognized by Comedi, and fills the pull-down lists for analog input and analog output automatically so that the user can choose which channels to use.  On exiting this panel, all choices are saved in a configuration file, so that the next time the program is started, these choices are loaded automatically.  The configuration file (indeed, all configuration files for the program) can be edited outside the program using a text editor.  One can independently choose which channel should be used for feedback.  This can be one of the input channels to monitor, or another channel entirely.  Normally, the feedback loop uses the value measured in the feedback channel as input to the PID algorithm.  For scanning modes where the feedback signal is a 
strong function of the distance between tip and surface (such as scanning tunneling microscopy), one can also choose logarithmic feedback.

The System Configuration Panel also allows one to choose a specific scanner.  Since piezoelectric scan tubes are generally hysteretic, the actual lateral and axial displacement of the scan tube is a function not only of the applied voltage, but also of the energization history.  If the hysteresis of the scan tubes is not taken into account, the images obtained will be distorted.  This problem can be eliminated by using closed loop scanners which use position sensors to determine the actual displacement of the tube.  However, implementing position sensors with their associated electronics is expensive.  An alternative method is to quantify the hysteresis by measurement, which can then be corrected for in software:  since the hysteresis is a function of the range, this should be done for specific scan ranges.  We have quantified the hysteresis in our scan tubes using a fiber-optic based intensity deflection sensor with submicron resolution (MTI-2100 Fotonic, \url{http://www.mtiinstruments.com}), and then used 
a third order polynomial to fit the 
deflection as a function of the applied voltage independently for voltage sweeps in opposite directions.  (Experimentally, we find that the deflection as a function of applied voltage is independent of the scan speed, at least at the very low sweep rates that are typically used for scanning.)  The calibration parameters for up-sweeps and down-sweeps for each scan range are stored in a scanner configuration file that is loaded when the program is started (or when the scanner is changed).  To obtain an image, the program steps evenly in distance, and so has to calculate at each step the voltage required to obtain the required deflection for that specific sweep direction by inverting the third order polynomial.  However, we have found that the parameters obtained in this way do not fully account for the distortion that is visible in the image.  Fine tuning of the calibration parameters can be done by scanning a standard sample and changing the parameters until the expected image is obtained, a capability that 
is built into recent versions of the open source SPM analysis software Gwyddion\cite{gwyddion} that is available for all platforms.  For variable temperature operation, this is likely the only method to obtain the calibration parameters, starting from the room temperature parameters scaled to account for the reduced scan range at lower temperatures. 

Once the system has been configured, the Coarse Approach Tool can be opened.  Coarse approach is achieved using the ``jog'' technique:  the scan tube is first fully retracted, and then slowly advanced step-by-step.  At each step, the program checks if the tip has approached the sample.  This state is defined by the  voltage read on the feedback channel being within a user-defined range of a user-defined setpoint.  If the tip has not approached the surface, the scan tube is fully retracted, the coarse approach mechanism brings the scan tube and tip one step closer, and the entire process is repeated again.  In order to ensure that the tip does not crash, the coarse step size should be less than the full extension of the scan tube.  However, in order to avoid a very large number of  ``jog'' steps, the coarse step size should be about half the full extension of the scan tube.    
In our case, the coarse step size is approximately 0.5 $\mu$m and the full extension of the scan tube is 2 $\mu$m.  Once the system has approached, one can take an ``acquisition'' curve; for example, for a tuning-fork AFM operated using a PLL, the acquisition curve is a plot of the frequency shift as a function of $z$, while for a scanning tunneling microscope, it would be the tunneling current as a function of $z$.  This curve can be saved for later analysis. 

After the tip has approached the sample, one can choose the required scanning mode:  here we discuss atomic force microscopy (AFM), although other modes (electrostatic force microscopy (EFM), magnetic force microscopy (MFM)) can also be implemented.  Prior to scanning, the system is brought under feedback, using a real-time PID control loop that is described in more detail below.  The input to the PID loop is read from the feedback channel defined in the System Configuration Panel:  this is subtracted from the setpoint to generate the error signal for the PID.  In terms of feedback control, the sign of the error signal for stable feedback is also important.  It can be different for different force transducers, or different points of stability if the approach curve is nonmonotonic.  In our program, the sign of the error signal can be set by multiplying the result of the subtraction by -1 if necessary.

The scan ranges are set by those defined in the scanner configuration file in the System Configuration Panel.  Except for the possibility of choosing line-by-line leveling and contrast adjustment of the image during the scan, no analysis is done in this program.  For such analyses, there are already a number of open source programs designed specifically for scanning probe microscopy. As mentioned above, we have used Gwyddion,\cite{gwyddion} which we find sufficiently powerful for our purposes.        
\begin{figure*}[!]
\vspace{0.25cm}
\includegraphics[width=15cm]{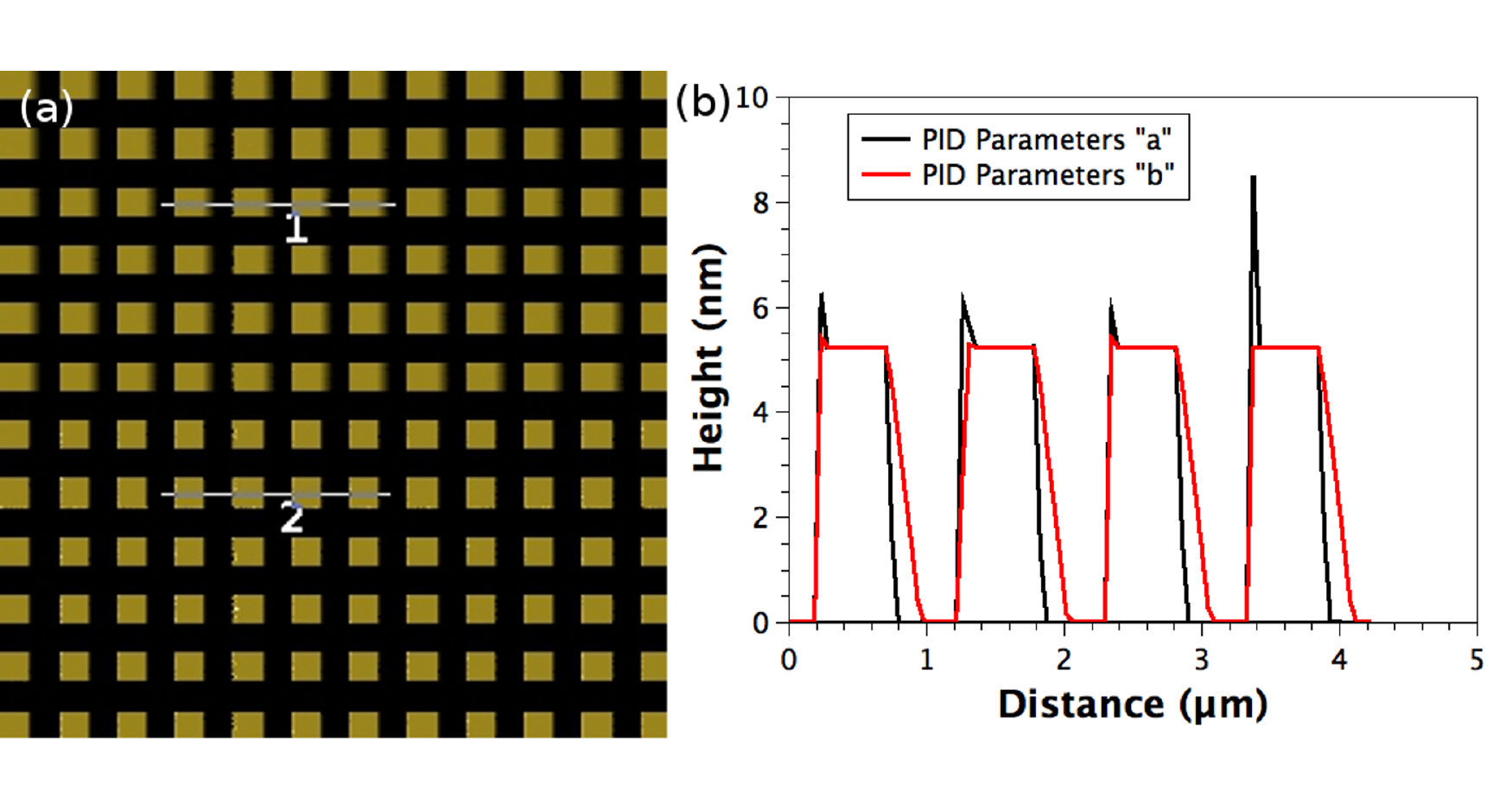}
\caption{Forward scan image of a simulated array of 5 nm high, 5 $\mu$m by 5 $\mu$m squares.  The parameters of the PID controlling the feedback loop were changed halfway through the scan. (a) Forward scanning image.  (b)  Line profiles corresponding to the two line sections labeled 1 and 2 in (a), corresponding to different PID parameters:  1: P=0.001, I=10 $\mu$s, D= 1 ns; 2: P=0.004, I=8 $\mu$s, D=1 ns.  }
\vspace{-0.5cm}
\label{fig4}
\end{figure*}
\subsection{Real-time control}
The main difference between RTSPM and other commercial and non-commercial programs is in the real-time feedback required for control.  In the context of the current program, this is primarily implemented in the feedback loop that controls the $z$ position of the scan piezo.  The feedback loop is a proportional-integral-differential (PID) type controller.  PID controllers are discussed extensively in the literature.  Our implementation is based on the incremental form from the text by \AA str\"{o}m and Murray.\cite{astrom}  The important section of our implementation of the PID loop is given below:\cite{github}  
\scriptsize
\indent
\begin{verbatim}
............

 bi = PropCoff*PIDLoop_Time/IntTime;   //integral gain
 ad = (DiffTIme)/(DiffTime+PID_cutoff_N*PIDLoop_Time);
 bd = PropCoff*PID_cutoff_N*PIDLoop_Time);

 Error = AmplifierGainSign*OutputPhase*(SetPoint - FeedbackReading);
 Pcontrib = PropCoff*(Error - LastError);
 Dcontrib = ad*LastDiffContrib - bd*(Error - 2*LastError +
                                               SecondLastError);
 v = LastOutput + Pcontrib + Icontrib + Dcontrib;

 //next, take care of saturation of the output....anti-windup
 PIDOutput = v;
 PIDOutput =(PIDOutput>MaxOutputVoltage)? MaxOutputVoltage:PIDOutput;
 PIDOutput =(PIDOutput<MinOutputVoltage)? MinOutputVoltage:PIDOutput;
      
 (.....output to card....)
      
 //Update the integral contribution after the loop
 Icontrib = bi*Error;

 //Update parameters
 LastError = Error;
 SecondLastError = LastError;
 LastDiffContrib = Dcontrib;
 LastOutput = PIDOutput;
 .............
\end{verbatim}
\normalsize
Here {\small\verb#PropCoff#}, {\small\verb#IntTime#} and {\small\verb#DiffTime#} are the PID parameters input by the user from the main program, while {\small\verb#PIDLoop_Time#} is the cycle time of the PID loop.  The parameters {\small\verb#bi#}, {\small\verb#ad#} and {\small\verb#bd#} are calculated only if these parameters have been changed.  {\small\verb#PID_cutoff_N#} is a parameter to limit the high frequency gain of the derivative term, and is set to 20 in our implementation.  The incremental nature of the algorithm means that it is bumpless, so that transitions into and out of feedback do not cause large changes in the $z$ position of the piezo, and includes so-called anti-windup so that the integral contribution does not overwhelm the control signal if the error signal going in to the algorithm is large for an extended period of time.  The correction term to the control signal is updated on each cycle of the PID loop.  Here one really benefits from the ``hard'' real-time nature of the control loop: 
apart from some jitter on the scale of 1-2 $\mu$s, the time between successive iterations of the PID loop (set by {\small\verb#PIDLoop_Time#}) is well determined.  As we have noted earlier, this time can be as short as 5-10 $\mu$s; however, at this rate, any program running on the main GUI freezes.  For our current setup, we have chosen a more conservative loop time of 50 $\mu$s.  With this choice, the jitter is a smaller fraction of the loop time, reducing the noise in the system.

Figure \ref{fig3} shows the response of the PID to a simulated square wave input measured on an oscilloscope.  The input was generated from a SRS 345 frequency generator (Stanford Research Systems, \url{http://www.thinksrs.com}); the control signal from the DAQ card was subtracted from this signal using a homemade instrumentation amplifier to generate the error signal input into the PID.  As the instrumentation amplifier (Analog Devices AD624) has a flat frequency response at unity gain out to 80 kHz, the response of the system is determined by the response of PID itself.  Figure \ref{fig3}a shows the response with the PID parameters optimized to obtain the quickest reponse with no overshoot.  The 50 $\mu$s loop time of the PID is clearly visible in the discrete steps of the response.  Increasing the proportional gain (Fig. \ref{fig3}b) results in overshoot, as expected.  For optimized parameters, the PID responds within $\sim$ 300 $\mu$s, which is sufficient for a 1-2 Hz scanning rate.   

\subsection{Imaging results}

Figure 4 shows an image scanned in the forward direction ($-x \rightarrow +x$) at a scan rate of 0.781 Hz (1.28 s/line).  The ``sample'' in this case is a real-time simulator program that simulates the response of a tuning-fork AFM, and will be described in detail elsewhere.  Briefly, the simulator takes as input a voltage from the RTSPM program that represents the height of the $z$ piezo, and outputs a voltage that represents the frequency shift of the tuning-fork, which is a function of the height of the tip above the sample.  The height in turn depends on the position of the tip on the ``sample.''  The simulator program determines the height of the sample by reading the voltages that the RTSPM outputs to control the $x$ and $y$ piezo directions, and comparing them with a predetermined pattern.  For the image of Fig. \ref{fig4}a, for example, if the $x$ and $y$ voltages are such that the scan position is within one of a regular array of squares of size 500 nm $\times$ 500 nm, the simulator program sets the 
height of the 
sample 
to be 5 nm, and outputs the voltage representing the frequency shift accordingly.

\begin{figure}
\includegraphics[width=8.5cm]{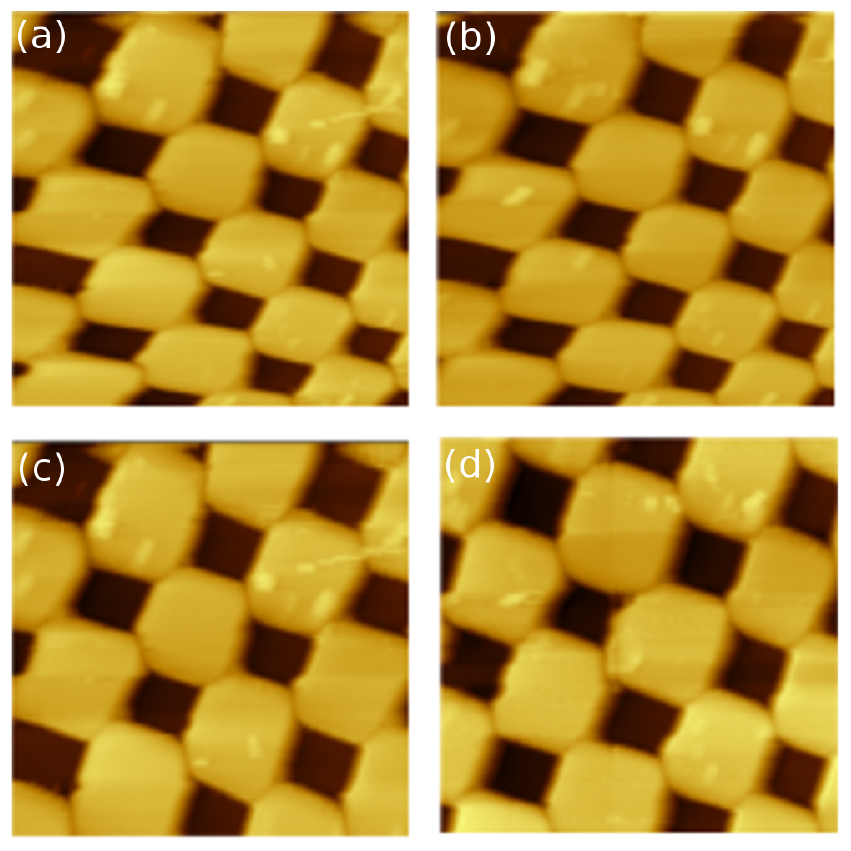}
\caption{Forward scan images of a standard square sample. The pitch is 10 $\mu$m. (a) Scan without any correction applied to compensate for the hysteresis of the scanner. (b) Scan of the same region as in (a) but with correction factors applied in the program from the MTI Fotonic sensor. (c) The image in (a) corrected for distortion using Gwyddion. (d) The image of the sample with the correction factors acquired from Gwyddion, but now implemented in the program while scanning. The correction factors are given in the text. }
\vspace{-0.5cm}
\label{fig5}
\end{figure}
Closer inspection of the image in Fig. \ref{fig4}a shows that in the top half of the image, the right hand edges of the squares are less clearly defined than the left hand edges; while in the bottom half the edges on both sides are sharp.  This is because the PID parameters of the feedback loop were changed halfway through the scan.  Figure \ref{fig4}b shows cross-sectional profiles corresponding to the two lines labelled ``1'' and ``2''  in Fig. \ref{fig4}a, with the corresponding PID parameters.  As is well known, overshoot and the resulting quality of the image are extremely sensitive to the chosen PID parameters.  At the moment, we adjust these parameters by hand, but implementing a simple tuning algorithm within the program should not be difficult.      

The simulator program does not take into account real experimental details such as the nonlinear, hysteretic behavior of the piezo scan tubes of a microscope.  To show the effect of these nonlinearities, Fig. 5a shows a scan of a standard topography sample (TGX 11P from Mikromasch, \url{http://www.spmtips.com/}) that consists of square depressions with a pitch of 10 $\mu$m.  The direction of the scan is in the forward direction:  the program scans lines from $-x \rightarrow +x$, with the $y$ position being stepped from  $+y \rightarrow -y$.  In Fig. \ref{fig5}a, no correction has been applied to take into account the nonlinearities of the piezo tube.  It is immediately apparent that these nonlinearities are significant:  the scan range in the lower right corner is much larger than the scan range in the upper left corner.   In Fig. \ref{fig5}b, a correction has been applied by fitting the displacement of the end of the scan tube measured by the MTI Fotonics optical sensor as a function of the voltage applied 
while sweeping in one direction.  The polynomial fit is 
of the form $x=A + B V + C V^2 + D V^3$, with a similar equation for the $y$ direction.  For this scan range, the fitting parameters were $A=-0.9658$,    $B=0.9042$,  $C=0.0162$, and $D= -0.000292$ for the $x$ direction, and $A=-1.1986$,    $B=1.1026$,  $C=0.0162$, and $D= -0.000292$ for the $y$ direction.  These parameters are then used in the scanner configuration file, with the program performing an interpolation to determine the voltages to generate a scan with fixed steps in $x$ and $y$.  Figure \ref{fig5}b shows the resulting scan.  While there is some improvement, there is still significant distortion in the scan.  

It appears that the correction factors that were generated by fitting the piezoelectric hysteresis curve measured by the Fotonic sensor are not correct.  In order to determine the correct parameters, we have used Gwyddion to determine by hand the parameters that result in the least distortion.  Figure \ref{fig5}c shows the result of applying this correction to the image in Fig. \ref{fig5}a, with parameters $A=-0.0$,    $B=1.1$,  $C=0.03$, and $D= -0.15$ for the $x$ direction, and $A=0.0$,    $B=0.9$,  $C=0.02$, and $D= -0.15$ for the $y$ direction.  These parameters significantly reduce the distortion.  Figure \ref{fig5}d shows another scan of the same sample with these new parameters in the scanner configuration file.  It is clear that the resulting image is much less distorted than either Fig. \ref{fig5}a or Fig. \ref{fig5}b, on par with the image in Fig. \ref{fig5}c.  We are not sure why the measurements with the Fotonics sensor give inaccurate results.  It may be due to the difficulty of making sure that 
one is really measuring the end of the tube, or that the measurement axes are truly orthogonal.  Nevertheless, the images in Fig \ref{fig5} indicate that using an image analysis program is a viable means of generating coefficients that can correct distortions in real time.  Indeed, given the difficulty of using an optical detector in extreme environments such as low temperatures or high magnetic fields, distortion correction using these techniques may be a more flexible alternative.

In summary, we have described scanning probe control software and hardware using a real-time operating system that significantly reduces the cost of building a scanning probe microscope.  The software is readily extensible to handle complicated scanning protocols if required.

\begin{acknowledgements}
This research was conducted with support from the US Department of Energy, Basic Energy Sciences, under grant number DE-FG02-06ER46346. 
\end{acknowledgements}

\end{document}